\documentclass{svproc}
\usepackage{color}
\usepackage{graphicx}        
\usepackage[bottom]{footmisc}
\usepackage{multicol}        
\usepackage{url}
\usepackage{booktabs}
\usepackage{float}
\usepackage{tabularx}
\usepackage{caption}
\usepackage[labelsep= space, labelfont = {sf, bf}, textfont = scriptsize]{caption}

\begin{document}
\definecolor{darkred}{rgb}{0.0, 0.0, 0.0}

\mainmatter              
\title{Mining the Automotive Industry: \\ A Network Analysis of Corporate Positioning and Technological Trends\thanks{Preprint Version}}
\titlerunning{Mining the Automotive Industry}  



\author{Niklas Stoehr\inst{1}\thanks{ORCID-ID: 0000-0003-2867-0236}, Fabian Braesemann\inst{2}\thanks{ORCID-ID: 0000-0002-7671-1920}, Michael Frommelt\inst{1}, Shi Zhou\inst{3}}
\authorrunning{Stoehr, Braesemann, Frommelt and Zhou} 
%
\tocauthor{Niklas Stoehr, Fabian Braesemann, Michael Frommelt and Shi Zhou}

\institute{IBM, AI Core, GER
\and
University of Oxford, Sa\"id Business School \& Oxford Internet Institute, UK\\
\and
University College London, Department of Computer Science, UK\\
\email{f.braesemann@sbs.ox.ac.uk}}

\maketitle              

\begin{abstract}
 The digital transformation is driving revolutionary innovations and new market entrants threaten established sectors of the economy such as the automotive industry. Following the need for monitoring shifting industries, we present a network-centred analysis of car manufacturer web pages. Solely exploiting publicly-available information, we construct large networks from web pages and hyperlinks. The network properties disclose the internal corporate positioning of the three largest automotive manufacturers, Toyota, Volkswagen and Hyundai with respect to innovative trends and their international outlook. We tag web pages concerned with topics like e-mobility \& environment or autonomous driving, and investigate their relevance in the network. 
 Sentiment analysis on individual web pages uncovers a relationship between page linking and use of positive language, particularly with respect to innovative trends. Web pages of the same country domain form clusters of different size in the network that reveal strong correlations with sales market orientation. Our approach maintains the web content's hierarchical structure imposed by the web page networks. It, thus, presents a method to reveal hierarchical structures of unstructured text content obtained from web scraping. It is highly transparent, reproducible and data driven, and could be used to gain complementary insights into innovative strategies of firms and competitive landscapes, which would not be detectable by the analysis of web content alone. 
\\

\textbf{Code and Data:} {\fontsize{8}{9.1} \url{www.github.com/Braesemann/MiningAutomotiveIndustry}}\\

\textbf{Version:} Preprint
\end{abstract}

\section{Introduction}\label{introdcution}

\subsection{Motivation}

Environmental change and the ongoing digitisation cause large-scale transformations in the economy, as boundaries of production, distribution and consumption are reshaped \cite{forum2017understanding,committee2017impact,forum2016digital}. Industries are impacted differently depending on factors like contribution to greenhouse gas emissions, automation capabilities, customer proximity, and labour complexity \cite{committee2017impact}. One of the biggest, yet most strongly affected industries is the automotive industry \cite{forum2016digital,oecd2011recent,knoedler2019automotive,gao2016disruptive,kuhnert2018five}.\\[1ex]
According to estimates of the International Organization of Motor Vehicle Manufacturers (OICA), more than 5\,\% of the world's total manufacturing employment is directly involved in the production of vehicles and parts \cite{oica2018world}. However, the economic importance of the automotive industry reaches far beyond that. Many of the most groundbreaking innovations of the 20th-century, mass production, just-in-time and multi-divisional business organisation originated in automotive companies and left a recognisable geographical footprint \cite{knoedler2019automotive,ferrazzi2011the}.\\[1ex]
The need for sustainable mobility solutions will drive further innovation in fields relevant to the automotive industry such as e-mobility, connectivity and autonomous driving \cite{braesemann_global_2019,stephany_coding_2019,stephany_exploration_2017}. These innovations provide unparalleled opportunities for value creation, but also major risks as new market participants might introduce services, which could threaten the established ecosystem \cite{knoedler2019automotive,gao2016disruptive}. Such radical shifts are creating the need for analytic feedback and status reports. Not only executive boards, policy-makers and investors, but more importantly, millions of employees are interested in a successful and sustainable transformation and therefore rely on objective assessments of car manufacturers' positioning with regards to novel developments and technologies.\\[-4ex]

\subsection{Contributions}

In this paper, we provide a new perspective on the possibilities for analysing shifting industries. Using only publicly available data, we present a network-centred approach that allows for conclusions on innovative orientation and international outlook of the world's leading car manufacturers, Toyota, Volkswagen, and Hyundai.\footnote{These are the three biggest automotive manufacturers by production numbers in 2016: Toyota (10.2 mio. vehicles), the Volkswagen Group (10.1 mio cars), and Hyundai with 7.9 mio. units \cite{oica2016world}.}
\\[1ex]
In this study, we enrich a quantitative analysis of web page content with meta-information. This approach surpasses conventional search engine optimisation and web page analysis, which are increasingly important tools for analysing markets and competitive landscapes, as for example, the recent introduction of the \emph{Google Market Finder} tool shows. Our method exploits the web page structure of the firms to obtain complex networks, where each web page is considered a node and each hyperlink referring to another web page is considered a link. This novel approach of exploiting the interplay between content and network analysis allows mapping a company's corporate positioning based on publicly available information. The approach is thus highly transparent and can be reproduced to monitor and contrast company web pages to gain insights into their innovation and market strategies.
\\[0.75ex]
The study exemplifies the value of applying network analysis as a complementary tool to text mining and content analysis. In times of abundant non-structured text information that can easily be obtained from different online sources via web-scraping, the combination of network analysis and content analysis allows for imposing structure on unstructured big data, as the hierarchy of information, which is implicitly stored in the web-page network topology, is maintained.

\section{Related Work}
\label{relatedwork}

This work is related to two strands of literature: studies on the transformation of the automotive industry and network analyses of web pages.\\[0.75ex]
Change in the automotive industry has been intensively investigated in past studies \cite{ferrazzi2011the,traub2018digitalization,fridman2018autonomous,gunther_role_2015,thoben2017industrie}, focusing on changing geographies and challenges of the technological transformation, and on technical, environmental, and management implications. Moreover, network analysis has been applied in the field, mainly to understand Supply-Chain systems in the automotive industry \cite{kito2014the,swaminathan2002network}.\\[1ex]
While web page networks of car manufacturers have, to our knowledge, not been investigated previously, similar approaches were applied in tourism and e-commerce. For instance, Baggio et al. investigate the links between tourism destination websites to find the statistical characteristics of the underlying graph \cite{baggio2007the}, and Wang et al. establish principles of an improved link structure for a hypothetical e-supermarket website \cite{wang2006optimal}.\\[0.75ex] 
Additionally, many pioneering studies in network science have focused on the world wide web: in 2000, Broder et al. discovered the bow tie structure of the world wide web \cite{broder2000graph}, and Meusel et al. aimed to understand it's growth mechanism in constructing a giant network of large parts of the world wide web \cite{meusel2014graph}. Barab\'asi and Albert used web data to demonstrate their theory on the emergence of scaling in random networks \cite{barabasi1999emergence}, and the PageRank algorithm was introduced for ranking web pages in a directed graph of the world wide web \cite{page1998the}.\\[0.75ex]
The content and structure of web pages has previously been investigated from the perspective of search engine optimisation \cite{ortiz-cordova2012classifying}, data retrieval~\cite{gowda2016clustering}, website design \cite{wan1998web}, and web navigation \cite{sahebi2008an}.\\[0.75ex]
Our approach builds on the previously mentioned studies, but assumes a different perspective. Instead of analysing content and structure of web pages to optimise properties of individual websites, we explore the potential of applying network analysis of web pages to gain insights into business models and innovative orientation of companies and industries.

\section{Research Hypotheses}

We analyse the properties of web page networks of leading car manufacturers and tag individual web pages that mention relevant keywords to observe their position in the network.\footnote{For this part of the analysis, which hinges on the identification of specific keywords, we analyse the US domains of the company websites ({\scriptsize \url{www.toyota.com}, \url{www.vw.com}, \url{www.hyundaiusa.com}}), as the United States is the second largest global car market behind China, and the largest English-language car market.} If car manufacturers aimed to associate themselves with innovative topics, which are transforming the automotive industry, it could be expected that this is reflected in the positioning of innovative keywords and their associated network centrality. This leads to our first research hypothesis:
\begin{quote}
$H1$\textit{: Car manufacturer web pages dealing with innovative topics tend to be higher ranked than other web content.}
\end{quote}
Furthermore, firms eager to display their innovative efforts in a good light should be more likely to describe important content with positive language as a means of customer communication. Accordingly, we hypothesise:
\begin{quote}
$H2A$\textit{: Well connected pages are more likely to show a positive sentiment than peripheral ones.\\
$H2B$: Pages focusing on innovative topics are characterised by a positive sentiment.}
\end{quote}
To test this, we perform sentiment analysis on the content of each web page in the network. Additionally, the web page hierarchy should allow to gain insights into the international orientation of the firms. This leads to our last hypothesis:
\begin{quote}
$H3$\textit{: The size of car manufacturer country domain networks corresponds to the country market size.}
\end{quote}
This hypothesis is examined via analysing the prevalence of different country domains and target markets throughout the network.\footnote{For the international comparison, we use the manufacturers' international web pages ({\scriptsize \url{www.toyota-global.com}, \url{www.volkswagen.com}, \url{www.hyundai.com/worldwide}}), as a starting point for the data collection.} 
\\[-4ex]


\section{Methodology}
\label{methodology}
\vspace{-1ex}

\begin{figure}[H]
\centering
\includegraphics[width = 1\linewidth]{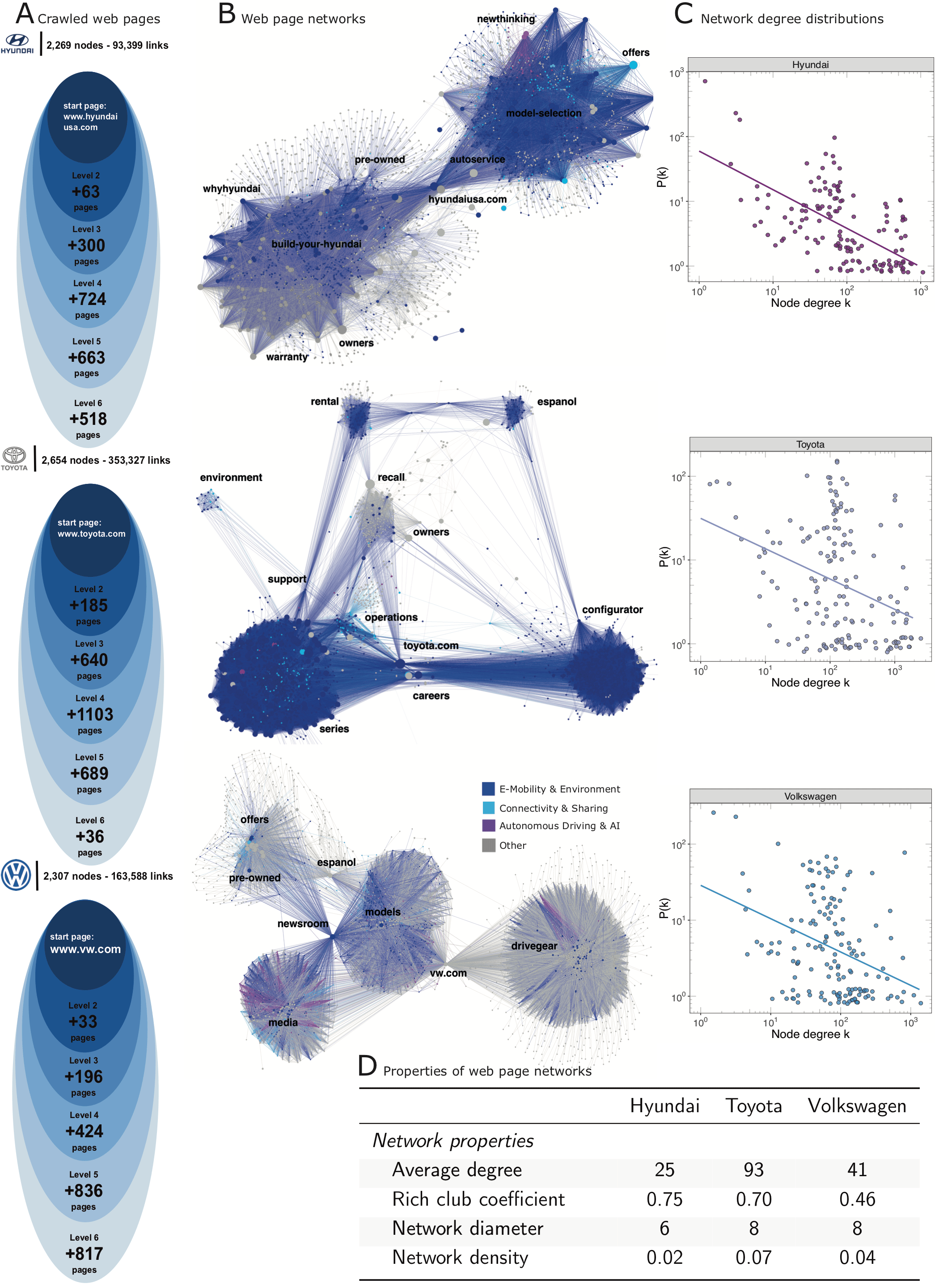}
\captionsetup{width=1\linewidth}
\caption{\footnotesize{\sf{(\textbf{A}) Crawled web pages of three car manufacturers: crawlers collected all sub-pages (nodes) and hyperlinks (links: in- and out-links are combined as undirected links) to a depth of 6 levels in a breadth-first manner. (\textbf{B}) Resulting web page networks: pages mentioning keywords on innovative trends are highlighted. The networks differ in their structure and positioning of innovative contents.  (\textbf{C}) Degree distribution of the networks (\textbf{D}) Network Properties: average node degree, average rich club coefficient, network diameter, and network density vary between the three manufacturer web pages.}}}
\label{fig:fig1}
\vspace*{-0.75em}
\end{figure}

Our approach consists of four steps. First, we apply web-crawling on the company web pages in order to obtain the network data. Secondly, we visualise the networks and derive properties and centrality measures from the data. Thirdly, we search for relevant keywords on innovative topics and apply sentiment analysis to the web pages. Lastly, we compare the manufacturers with respect to their international orientation in 25 national markets.\\[1ex] 
After retrieving the network data, we turn towards a comprehensive analysis of link structure and node meta-information, using a number of tools. We use the graph visualisation tools \textit{Gephi} \cite{bastian2009gephi} and \textit{Graphviz} \cite{ellson2003graphviz}. The Python package \textit{NetworkX} \cite{hagberg2008exploring} is used for obtaining the network and node properties. The search engine optimisation software \textit{Screaming Frog} and \textit{SEO Powersuite} are employed to get additional node information.\\[1ex] 
To obtain the network data, we implement a crawler able to retrieve an extensive number of pages from the web. For this purpose, we use the \textit{Python} package \textit{Beautiful Soup} for HTML and XML parsing. Essentially, the crawler starts at a single initial web page, where it retrieves the web page including all it's hyperlinks. Following the breadth-first paradigm, it then visits all the web pages that the start page links to and stores them as nodes connected via the hyperlinks (links). The crawler then repeats this process on the second level of web pages before it goes on to the next level. We terminate the crawler after the hyperlinks on the 6th level have been added to the network (Fig.\,\ref{fig:fig1}A). The data was collected in December 2018.\\[1ex] 
Based on the key themes identified in the literature \cite{forum2016digital,oecd2011recent,knoedler2019automotive,gao2016disruptive,kuhnert2018five}, we establish three major innovative trends affecting the automotive industry: (1) 'e-mobility and environment', (2) 'autonomous driving and artificial intelligence', and (3) 'connectivity and shared mobility'. These trends are identified in the textual components\footnote{The textual components are all HTML tags (predominantly "title" and "body") of a web page. Specifically excluded are the tags "script", "style", "head", "[document]". This way, we only include textual components visible to the user.} of each web page by counting relevant keywords.%
\footnote{The list of keywords has been created prior to looking at any company website, based only on the qualitative definitions in the literature. The respective keywords are:%
\vspace{-1ex}
\begin{description}
\item[E-mobility \& environment:]e-mobility, battery, environment, biological, eco, ecological, electric, hybrid, environment, environmental-friendly;
\item[Connectivity \& shared mobility:]connectivity, shared, mobility, sharing, interconnectedness, cloud, cloud computing, wifi, 5G;
\item[Autonomous driving \& artificial intelligence:]autonomous, self-driving, ai, machine learning, artificial intelligence, intelligent, neural network, algorithm.
\end{description}} 
\\[1ex]
For the sentiment analysis, we use the natural language processing toolbox \textit{TextBlob} \cite{loria2018textblob} and analyse the textual components of each individual web page with regards to their polarity on a continuous, symmetric sentiment scale ranging from -1 (negative) to +1 (positive) \cite{falck2019measuring}.\\[1ex] 
To capture the international orientation of the car manufacturers, we examine the peculiarity and prevalence of web pages of different country domains, starting from the international landing page of the three firms. The domain affiliation is derived from the country tag, e.g '.ca' for Canada, '.uk' for United Kingdom etc. The number of web pages per country domain are then compared with the size of the national market in terms of car sales and registrations \cite{oica2017markets}.\\[1ex]
At this point, we want to emphasise that this study is exploratory: the purpose is to investigate the value of complementing content analysis with network centrality measures on the example of car manufacturer web pages.

%
%

\begin{figure}[H]
\centering
\includegraphics[width = 1\linewidth]{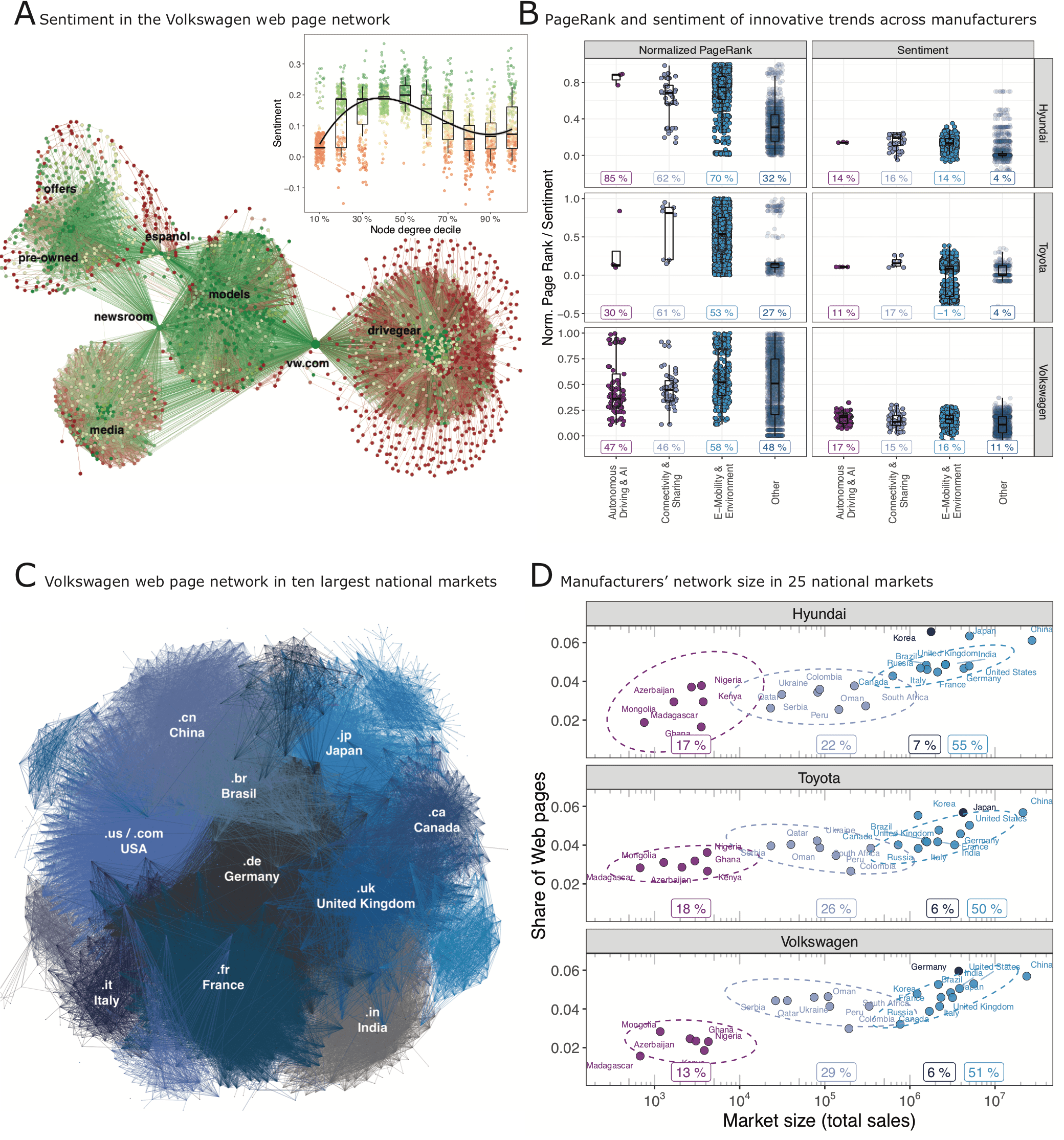}
\captionsetup{width=1\linewidth}
\caption{\footnotesize{\sf{(\textbf{A}) Distribution of sentiment in the Volkswagen web page network: nodes with intermediate degree centrality show, on average, a more positive sentiment (inset). (\textbf{B}) Distribution of PageRank centrality and sentiment per topic and manufacturer: pages on innovative topics (in particular e-mobility and environment) tend to be higher ranked and to have more positive sentiment than other sub-pages. (\textbf{C}) International Volkswagen web page network (crawled from international landing page) in ten largest car markets: Volkswagen's German home market network forms a central cluster, densely connected with other key key markets in Europe and North America. (\textbf{D}) Global sales (log-scale) and share of web pages in market groups: the three manufacturers similarly focus their web presence on a cluster of large car markets, with their home market web page network being disproportionately extensive.}}}
\label{fig:fig3}
\vspace*{-1.25em}
\end{figure}

\section{Results}
\label{results}

\subsection{Network Structure}

The obtained web page networks of the US domains of Hyundai, Toyota, and Volkswagen show distinct structures (Fig.\,\ref{fig:fig1}B). In the Toyota and Volkswagen networks, several navigational pages connect the major elements of the websites, while the web pages of Hyundai fan out into only two major components.\\[1.5ex]
The networks also differ with regards to the positioning of innovative topics.\footnote{In the network visualisations, the nodes are coloured according to the keyword-category that appears most often in a web page. If none of the keywords occurs, the node is coloured in grey.} In contrast to Volkswagen, Hyundai and Toyota refer to 'e-mobility \& environment' in large parts of their networks; also in the sections 'model-selection' and 'configurator'. This result most likely reflects their focus on hybrid vehicles. All manufacturers refer to 'connectivity \& sharing' only in peripheral categories ('offers' and 'operations'). The topic 'autonomous driving \& artificial intelligence' is discussed in designated sections on Hyundai's 'newthinking' and Volkswagen's 'media' website; in contrast, it is essentially absent from Toyota's web page. These results provide support for hypothesis $H1$: the manufacturers tend to display content on innovative topics at prominent places in their web page networks\\[1ex]
Despite these differences, the networks show a similar degree distribution and network properties (Fig.\,\ref{fig:fig1}C and Fig.\,\ref{fig:fig1}D). According to the high nodal average degree of 93 and a network density of 0.07, the web page network of Toyota with its 2,654 nodes and 353,327 links is the largest and most densely connected one.

\subsection{Sentiment and Centrality}

To understand the role of positive language on the manufacturer web pages ($H2$), we display the Volkswagen network coloured according to sentiment (Fig.\,\ref{fig:fig3}A). Qualitatively, we note a positive correlation between node degree and sentiment: more central nodes appear to have a more positive sentiment. As the inset shows, node degree centrality and sentiment are actually characterised by an inverse U-shaped relation. In contrast to our initial hypothesis $H2A$, not the most central pages, but intermediately ranked pages reveal the most positive sentiment. These pages are most likely content-driven (e.\,g. pages that are linked to in the 'media' or 'newsroom' sections of the website). Less central nodes might rather describe specific topics such as warranty issues or model specifications, and the most central nodes appear to fulfil navigational purposes, characterised by a more neutral sentiment. Awareness of this arguably unintentional but yet noticeable use of language can provide a competitive edge, when firms consistently associate their content on innovative topics with positive sentiment.\\[1.5ex]
In all three manufacturer web page networks, sites dealing with innovative content tend to be more central than other web pages (Fig.\,\ref{fig:fig3}B, left panel); providing quantitative support for $H1$. The figure displays the distribution of the normalised PageRank per page (PageRank divided by the highest ranked page's value) and the average rank per category (labels below the box plots). Pages with the theme 'e-mobility \& environment' are, on average, more central than 'other' web pages with a normalised PageRank of 0.7 vs. 0.32 (Hyundai), 0.53 vs. 0.27 (Toyota), and 0.57 vs. 0.48 (Volkswagen).\\[1ex]
With regards to hypothesis $H2B$, sites on innovative themes use more positive language (Fig.\,\ref{fig:fig3}B, right panel). An exception is Toyota's content on e-mobility with a slightly negative average sentiment of -0.01. We could not find statistical significant relations between page centrality and sentiment per topic category. Nonetheless, the results suggest that car manufacturers tend to display content on innovative trends at more prominent positions in their networks. These findings emphasise the importance of linking content analysis with structural properties that maintain the hierarchical structure of the analysed content. Considering the web page network allows for identifying the manufacturers focus on certain topics. Without considering the network, the firms prioritisation of these topics would have not been detectable, given the relative low count of pages with innovative content (in particular on 'AI' and 'Connectivity' related topics).
\vspace{-2ex}
\subsection{Country Domain Analysis}

For analysing the international orientation of the firms, we crawl web page network data from the firms international landing pages. Figure\,\ref{fig:fig3}C shows the Volkswagen network of the sub-domains referring to the ten largest national markets in terms of car sales and registrations. The country domains agglomerate in densely connected clusters of different size, indicating the importance of the respective target markets. Volkswagen's German web page cluster is disproportionately large and connected with most other markets. Hence, it is at the core of the international network, reflecting Volkswagen's focus on it's home market.\\[1.5ex]
This finding does not only hold for Volkswagen, but also for the other two car manufacturers (Fig\,\ref{fig:fig3}D). The figure shows the size of the sub-domain networks (number of web pages) per manufacturer in 25 national markets for which we could obtain the number of passenger car sales \cite{oica2017markets}. Applying a k-means clustering algorithm on the number of sales (log-scale) yields three distinct groups of national markets: small markets of African and Asian countries (less than 1,000 cars sold in 2017), a second cluster of Global South countries with an intermediate market size (1,000\,--\,1,000,000 cars), and group of twelve large markets (more than 1 million cars), consisting of OECD and Newly Industrialised countries.\\[1ex]
As hypothesised ($H3$), the manufacturers show a similar pattern in terms of their web presence in these groups: 13\,--\,18\,\% of the firms' web pages target the group of small markets, while 22\,--\,29\,\% relate to the medium-sized markets. With 50\,--\,55\,\% of all their web pages, the firms' online presence clearly focuses on the group of large national markets with more than a million annually sold cars. The home market's web page network of each company is disproportionately large with 6\,--\,7\,\% of all web pages. This finding reveals the companies' competitive strategies of online marketing activities and resource allocation.
\vspace{-5ex}
\section{Discussion}
\label{discussion}
\vspace{-1.5ex}
\subsection{Summary}
\vspace{-1ex}
In the digital era, the automotive industry remains one of the cornerstones of global manufacturing, not only in terms of employment and trade, but more importantly for its role in introducing new technologies\cite{ferrazzi2011the}. This work presents a network-centred approach to gain insights into the innovative focus of three large car manufacturers by analysing the firms' web pages. In crawling publicly available content and structural meta-information from the websites of Toyota, Hyundai, and Volkswagen, we construct complex networks and analyse their properties. The interplay of content and hierarchical meta-information reveals meaningful patterns hidden from conventional web page content analysis. 
\\[1ex]
The analysis of the web page networks reveals that the three firms centrally present content on the topic 'e-mobility \& environment'; however, the topics 'connectivity \& sharing' and 'autonomous driving \& artificial intelligence' are dealt with only in more specialised sections. The companies tend to make use of more positive language on pages on innovative trends and emphasise such content at prominent places of the networks. Our analysis of national sub-domains shows that the manufacturers concentrate their online marketing efforts on a limited number of target markets, with a particular focus on their home market. Our approach exploits only publicly available information and is easily reproducible. It promotes transparency and could be used as a complementary tool to monitor shifting industries and competitive landscapes, as it helps to discriminate unstructured textual web data by relevance, which is implicitly captured by the hierarchical web page network structure.\\[-5ex]

\subsection{Limitations and Future Research}

Since this is an exploratory study, it has some limitations and future research is needed to establish generalizability and applicability of our findings beyond the case presented. 
Web-crawling in an iterative manner can yield instable results, depending on the starting page of the crawler. Thus, data collection should comprise an ensemble of methods conducted from several starting pages. Moreover, websites change frequently; hence, our approach is time-sensitive. To leverage it's potential, the websites should be monitored over longer periods of time, so that evolving structures and topics can be identified \cite{stoehr2019disentangling}.\footnote{Alternatively, the web page structures of those countries could be compared, in which the manufacturers apply different marketing strategies (there might be markets where the adoption rates of digital technologies in the automotive sector are higher).}\\[1ex]
Moreover, mentioning digital trends does not necessarily accompany putting the digital transformation into practice, but it demonstrates awareness and prioritisation. If many highly ranked pages are concerned with digital trends, the website is more likely to be presented in a search engine's result when a user searches for "e-mobility", "autonomous driving", or other innovative trends.

%
%

\end{document}